\documentclass[11pt,a4paper]{article}
\usepackage[T1]{fontenc}
\usepackage[utf8]{inputenc}
\usepackage{authblk}
\usepackage{longtable}
\usepackage{url}
\usepackage[ruled]{algorithm2e}
\usepackage{multirow} 
\usepackage{wasysym }
\usepackage{graphicx}
\usepackage[margin=1.2in]{geometry}

\title{Organization Mining Using Online Social Networks}
\author{Michael Fire, Rami Puzis, and Yuval Elovici\thanks{Email:\{mickyfi,puzis,elovici\}@bgu.ac.il}}

\affil{Telekom Innovation Laboratories  at Ben-Gurion University of the Negev \\
Department of Information Systems Engineering, Ben-Gurion University}

\date{}

\begin{document}
\maketitle
% Page heads

\begin{abstract}
Mature social networking services are one of the greatest assets of today's organizations.
This valuable asset, however,  can also be a threat to an organization's confidentiality. Members of social networking websites expose not only their personal information, but also details about the organizations for which they work. 
In this paper we analyze several commercial organizations by mining data which their employees have exposed on Facebook, LinkedIn, and other publicly available sources. 
Using a web crawler designed for this purpose, we extract a network of informal social relationships among employees of a given target organization.  
Our results, obtained using centrality analysis and Machine Learning techniques applied to the structure of the informal relationships network, show that it is possible to identify leadership roles within the organization solely by this means. 
It is also possible to gain valuable non-trivial insights on an organization's structure by clustering its social network and gathering publicly available information on the employees within each cluster. 
Organizations wanting to conceal their internal structure, identity of leaders, location and specialization of branches offices, etc., must enforce strict policies to control the use of social media by their employees.
\\\\
\noindent \textbf{Keywords.} Organizational data mining, Social network data mining, Social networks privacy, Organizational social network privacy, Facebook, LinkedIn, Machine learning, Leadership roles detection

\end{abstract}

\section{Introduction}

In recent years, online social networks have grown in scale and variety and today offer individuals the opportunity to publicly present themselves, exchange ideas with friends or colleagues, and network more widely.
For example, the Facebook\footnote{\url{http://www.facebook.com}} social network has more than 1.11 billion monthly active users, with new users signing up each month. According to recent statistics published by Facebook, on average 655 million Facebook users log onto this site on a daily basis, and more than 4.75 billion pieces of content are shared each day (web links, news stories, blog posts, notes, photo albums, etc.)~\cite{fbtechcrunch}.
On the one hand, social networks create new opportunities to develop friendships, share ideas, and conduct business.
On the other hand, many social network users expose personal third-party details about themselves and their social connections via their profile pages~\cite{acquisti,boshmaf}, as well as sensitive business information and details about their place of employment. 

In this study, we analyze publicly available social network data in order to infer the internal organizational structure of six high-tech companies of different scales. A similar analysis has been performed by Tyler et al. on the Hewlett-Packard  organization~\cite{tyler}. 
However, their analysis was based on protected organizational data, i.e., email logs.  
We show that it is possible to use only publicly available data, such as from Facebook and LinkedIn,\footnote{http://www.linkedin.com} in order to achieve similar results for multiple organizations. 

The contributions of this paper are threefold: First, we present a method for uncovering an organization's informal social network topology based solely on publicly available data. 
Second, we use the organization's structure to discover hidden leadership roles within the organization and to identify communities  inside the organization. 
Lastly, we perform a qualitative analysis of these leadership roles and communities  and demonstrate that it is possible to obtain significant insights into the organization and the role of each community without having any access whatsoever to the organization's internal data.

\subsection{Our Approach in a Nutshell}
The organizational mining methods proposed in this paper were applied to six well-known high-tech companies of various sizes, ranging from small companies with several hundred employees to large-scale companies with hundreds of thousands of employees.
For each company, the mining process included three major steps. 
First, we acquired the organization's informal social network topology from publicly available information,  as detailed in Section~\ref{crawler}.
As part of this process, we collected information about the company's structure as exposed by the company's employees on Facebook.
The presented method for organizational data mining can yield a wide range of organization social network topologies which were not available to the research community in the past. 

Next, we used different centrality measures to detect the hidden leadership roles inside each organization. 
In Section~\ref{managers}, we highlight the centrality measures with the highest accuracy in pinpointing the leaders.
We additionally used Machine Learning algorithms to classify management roles in each organization.

In the third step, we used a state-of-the-art algorithm to cluster the organization's social network into disjoint communities, and we cross-referenced the disclosed leaders and communities with information obtained from LinkedIn (see Section~\ref{communities}).
This enabled us to derive the roles of many communities within an organization, providing important insights about the organization. 
Such insights included, for example, the geographic deployment of the organization, the structure of the organization's different  divisions, the relationships between divisions, companies that were previously acquired, and the research focus of the organization.
These details can help us better understand both the structure and the communication patterns within organization. 
They also highlight the need for organizations to be aware of their social networking vulnerability and to establish policies to control this exposure as necessary.

The remainder of this paper is organized as follows. 
In Section~\ref{background}, we provide a brief overview of previous relevant studies on social network analysis with a special focus on organizational social network analysis. 
In Section~\ref{crawler}, we describe the methods used to obtain the organizational social network structure, and we show the different organizational datasets obtained. 
In Section~\ref{managers}, we present methods for identifying an organization's leadership roles. 
Next, our methods used to discover the communities' roles inside each organization are described in Section~\ref{communities}.
Lastly, in Section~\ref{conclusions}, we present our conclusions and offer future research directions.

\section{Background}
\label{background}
In this section, we describe previous work in the fields of online social networks and organizational social networks. 
We also provide an overview of studies that have used different types of data to reveal informal connections among an organization's employees in order to discover the organization's social network.

\subsection{Online Social Networks}
In recent years, the use of online social networks has grown exponentially.
Online social networks such as Facebook, Twitter,\footnote{http://www.twitter.com} LinkedIn, Flickr,\footnote{http://www.flickr.com} and YouTube\footnote{http://www.youtube.com} serve millions of users on a daily basis. With this increased usage, new privacy concerns have been raised. These concerns result from the fact that online social network users frequently publish information about themselves and their work-places. In 2007, a study carried out by Dwyer et al.~\cite{dwyer}  determined that 100\% of people who participated in the study had used real names on their Facebook accounts and 98.6\% had added photographs of themselves to their Facebook accounts. Moreover, in 2011, Boshmaf et al.~\cite{boshmaf} collected and analyzed more than 250GB of Facebook users' data and evaluated the amount of personal information  exposed by users. They concluded that many Facebook users disclose detailed personal information, including date of birth, place of work, email address, relationship status, and phone number. By using publicly available data from Facebook and cross-referencing it with other public data sources on the web, such as Google\footnote{http://www.google.com} and LinkedIn, one can infer further details about a Facebook user, such as specific work experience and areas of expertise. For example, Pipl\footnote{http://pipl.com} and PeekYou\footnote{http://www.peekyou.com} are able to search for information about a person across different social networks. These people search engines aggregate the obtained results and present a fully detailed personal profile. 

In this study we used publicly available data from Facebook in order to identify which Facebook users worked for a specific organization. We then cross-referenced the users' details with LinkedIn, Google search results, and the company's own web page in order to reveal the users' positions in the organization.

\subsection{Organizational Social Networks}
In the past six decades, a considerable amount of research has gone into analyzing and understanding communication patterns between individuals inside organizations. In 1951, Jacobson and Seashore~\cite{jacobson1951} were among the first researchers to study communication patterns among federal agency employees. 
In 1968, Pugh et al.~\cite{pugh1968} studied five primary dimensions of organizational structure applied to  52 different organizations in England.
In 1969, Allen and Choen~\cite{allen} studied technical communication patterns and their influences within two research and development laboratories at MIT. 
In 1979, Tichy et al.~\cite{tichy} presented a method for analyzing organizations using a network framework which included many network structural properties, such as centrality, clustering, and density. 
Tichy et al. used this framework to perform a comparative analysis of two organizations with several hundred employees. 
In 1991, Sparrow~\cite{sparrow1991} presented a method for using social network structural analysis to better understand criminal organizations. 
In 2002, after the tragic events of September 11, 2001, Krebs~\cite{krebs2002mapping} studied Al-Qaeda's organizational network structural properties and succeeded in identifying the conspiracy leader by using the degree and closeness structural properties of vertices. 
In 2003, Campbell et al.~\cite{campbell2003expertise} presented algorithms for expertise identification using email communication patterns. 
Their algorithms were evaluated on two different organizations. In our  own study, we show that expertise, leadership, and the roles of communities can be identified using publicly available data sources even without having access to internal organization data, such as email logs. 
Since 2004, after the release of about 500,000 Enron employees' emails~\cite{shetty2004enron}, many researchers have utilized this internal email dataset to better understand the Enron corporation's social network  and to discover various insights about the organization~\cite{diehl2007relationship, diesner2005communication,mccallum2005topic,shetty2005discovering, wilson2009discovery}.

In recent years with the increasing prevalence  of online social network usage, many studies have addressed the use and benefits of both public and internal social networking services to organizations.  
In 2009, Steinfield et al.~\cite{steinfield2009bowling} studied the connection between social capital and the use of social networking services deployed inside organizations. In the same year, Rooksby et al. published a detailed report on how online social networks are used in the context of the workplace~\cite{rooksby2009social}. Comprehensive reviews on organizational social networks can provide further insights~\cite{kilduff2003social,provan2007interorganizational,kilduff2010organizational}.

\subsection{Discovering an Organization's Social Network from Informal Connections}
The work reported in this paper is closely related to a 2004 internal study on the Hewlett-Packard organization carried out by Tyler et al.~\cite{tyler}.
By analyzing the organization's email corpus, which contained more than one million messages, they discovered the organizational social network topology and identified communities inside the organization. 
The authors used the betweenness-centrality measure~\cite{freeman1977set} to detect leadership roles within the organization. 
They also applied a version of the Wilkinson and Huberman algorithm~\cite{wilkinson2004method} which partitions the organization's social network into communities.
The results were  evaluated by interviewing several employees about the community they were automatically  placed in by the community detection algorithm. 
Naddaf and Mutyala~\cite{naddafa} presented a similar study in 2010.
They demonstrated a method for extracting informal social networks formed by employees of an organization based on the employees' email records.
They tested their method on a large public sector client and identified the authority of the employees by using the PageRank measure~\cite{page1999pagerank}. 
Moreover, Naddaf and Mutyala used the Fast Modularity algorithm~\cite{clauset2004finding} to identify communities in their client's organizational social network.

\section{Organization Social Network Crawler}
\label{crawler}
Many different types of web crawlers have been developed to collect data from large scale online social networks~\cite{mislove2007socialnetworks,boshmaf,gjoka2011multigraph,fire2012link}. 
Social networks crawlers usually start from several seed profiles and gradually expand the set of acquired profiles using, for example, Breadth-First-Search (BFS) crawling or other methods, such as Random-Walks~\cite{gjoka2011multigraph}.  

Unfortunately, standard social network crawling techniques are insufficient for performing data collection which focuses on a specific organization. 
During a preliminary study performed using BFS crawling, we collected many irrelevant profiles and often skipped Facebook users who worked in our targeted organization. 
To tackle the problem of targeted acquisition of profiles from online social networks, we developed an organization crawler which optimizes data collection from users associated with a specific group or organization.  
Our organization crawler utilizes the homophily principle~\cite{mcpherson2001birds}. According to the homophily principle, it is  more likely that a person has been employed by a certain organization if many of his or her friends have been employed by the same organization as well. 

In order to mine the social network for the profiles of employees from a selected target organization, our crawler worked according to the algorithm depicted in Algorithm~\ref{crawler1}.
The crawl starts from a set of seed profile pages initially identified as belonging to employees of the targeted organization.
The initial set of seeds can be obtained using a search engine. 
These seeds are used to initialize a priority queue (line 2).
All seeds have an initial priority of zero. 
Later, the priority of profile pages in the queue is increased with every friend that is employed by the target organization (lines 11-12). 

We proceeded by iteratively processing the next profile page with the highest priority (i.e., likelihood of being an employee of the target organization) until no potentially valuable profiles were left in the priority queue (lines 4-5). 
Every processed profile page was downloaded (line 7) and automatically analyzed. 
We employed a heuristic in an attempt to discover whether or not the currently processed profile page belonged to an employee of the target organization. 
This heuristic matched various keywords associated with the organization to the semi-structured data that appears in the user's publicly available Facebook profile. 
For example, in order to identify users from Ben-Gurion University's Information System Engineering Department, the crawler searched for strings such as ``Ben-Gurion Information System Engineering,'' ``BGU ISE,'' or ``ISE BGU'' in the collected profile page. 
In case we did not find a match to any of the keywords, we continued on to the next profile in the priority queue.
If, however, we did find that the dequeued profile page belonged to an individual who worked in the targeted organization, we collected the list of his or her Facebook friends (lines 8-9). 

Profile pages of Facebook friends that were already processed were ignored (line 10).
We increased the priorities of all friends waiting in the priority queue (lines 11-13). 
Afterwards, we inserted all newly encountered Facebook friends of the currently processed profile into the priority queue, with a priority of one (lines 14-16). 
This process repeated with the next profile page extracted according to the updated priorities. 

The crawler whose pseudo code is described by Algorithm \ref{crawler1} stops when the queue is empty. 
We will refer to this crawler as Version 1.
We also evaluated an optimized version of the organization crawler. 
This version tracked the number of friends within the targeted organization for each user profile in the priority queue and also the number of organization employees discovered during the last iterations. 
We stopped the crawling process if all users in the priority queue had at most one friend in the targeted organization and if the last thousand profiles acquired from Facebook did not belong to those of the organization's employees.
We will refer to the crawler with this stricter stopping condition as Version 2. 

\begin{algorithm}[t]
\SetAlgoNoLine
\KwIn{A set of seed URLs (S) to Facebook profile pages of organization's employees.\\
 A set of crawling organization target names, N.}
\KwOut{A set of Facebook profiles and their connections.}
$Q \leftarrow$ Priority-Queue() \\
$\forall_{URL \in S},Q.Enqueue(URL : 1)$ \\
$Crawled \leftarrow \invdiameter $\\
\While{$(Q \neq \invdiameter)$}{
$URL \leftarrow  Q.Dequeue()$ \\
$Crawled \leftarrow Crawled \cup \{URL\}$\\
$Page \leftarrow DownloadProfilePage(URL)$\\
\If{ Page contains N}
{
  F\_URLs $\leftarrow$ Extract list of friends from Page\\
  F\_URLs $\leftarrow$ F\_URLs $-$ Crawled\\
  \For{( F\_URL$\in$ F\_URLs$\cap$Q  )}{
	Increase priority (Q, F\_URL)  
  }
  \For{( F\_URL$\in$ (F\_URLs$-$Q)  )}{
	Q.Enqueue(F\_URL:1)  
  }
}
}
\Return Collected pages

\caption{Organization Social Network Crawler (Version 1)}
\label{crawler1}
\end{algorithm}

\subsection{Ethical Considerations}
During this study, we used our organization crawlers to collect a considerable amount of data from public sources regarding the studied organizations and their employees. To the best of our knowledge, Ben-Gurion University of the Negev regulations do not require explicit approval by an ethics committee for studies that involve publicly collected data. 
Nevertheless, in order to protect the privacy of the organizations' employees and the discovered confidential details of the organizations, we anonymized the organizations' names throughout this paper. Additionally, in the attached published datasets, we anonymized the employees' Facebook identities by randomly replacing the users' Facebook IDs  with a series of contiguous integers.

\subsection{Collected Organization Datasets}
\label{datasets}
In order to test the methods of organization data collection reported in Section~\ref{crawler}, we used our organization social network crawler to collect publicly available data from six commonly known high-tech companies.

The organization crawling results are depicted in Table~\ref{crawler_table}, where all the organizations' data were obtained during 2012. 

We grouped the companies based on their size:
Small (S), currently employing 500 to 2,000; Medium (M), employing 4,000 to 20,000; and Large (L) having more than 50,000 employees. 
Data on one company of each scale was acquired using each version of the organization crawler. 
We refer to the three companies targeted by Version 1 and Version 2 of the crawler as S1, M1, L1; and S2, M2, and L2, respectively. 

In the following subsections, we describe in detail the properties of each collected organization dataset (see Table~\ref{datasets_table}).
We used Cytoscape~\cite{shannon2003cytoscape} software to visualize the social networks formed by the employees of each organization.\footnote{All the organizations' graphs presented throughout this paper are embedded as Scalable Vector Graphics (SVG) image, which enables the reader to zoom in and view each node in each graph.}
The vertex colors in Figures~\ref{S1}-\ref{L2} represent various cluster roles, as will be explained in Section~\ref{communities}.
The analysis results of these networks are reported in Sections~\ref{managers} and~\ref{communities}.

\begin{table}[ht] 
\begin{center}
\caption{Organization Crawling Results\label{crawler_table}}{ % title of Table 
\centering
\begin{tabular}{c | c c c c c} % centered columns (4 columns) 

\hline %inserts double horizontal lines 
  \textbf{Org. }     &	\textbf{Crawler} &	\textbf{\#Total Crawled	}& \textbf{\#Org. Crawled}  & 	\textbf{Precision }\\  % inserts table heading 
            & \textbf{Version }&  \textbf{Profiles}& \textbf{Profiles}        &             \\
\hline
 S1 & Version 1	&22,992	&165	  &0.7\% &\\
 S2 & Version 2	&3,312  &320    &9.6\% &\\
 M1 & Version 1	&11,247	&1,429	&12.7\% &\\
 M2 & Version 2	&7,422	&3,862	&52.0\% &\\
 L1 & Version 1	&13,505 &5,793	&42.9\% &\\
 L2 & Version 2	&18,810	&5,524	&29.3\% &\\
\hline % inserts single horizontal line 
\textbf{Total}& \textbf{-}	&\textbf{77,288}& \textbf{17,096} &\textbf{22.1\%} &\\
\hline %inserts single line 
\end{tabular} 
}
\end{center}
\end{table}

%\begin{figure}[!t]
%\centering
%\includegraphics[width=2.5in]{myfigure}
% where an .eps filename suffix will be assumed under latex, 
% and a .pdf suffix will be assumed for pdflatex; or what has been declared
% via \DeclareGraphicsExtensions.
%\caption{Simulation Results}
%\label{fig_sim}
%\end{figure}

\subsubsection{Small Hardware Company (S1)}
\begin{figure}[t]
\begin{center}
\includegraphics[
width=0.8\textwidth,clip]{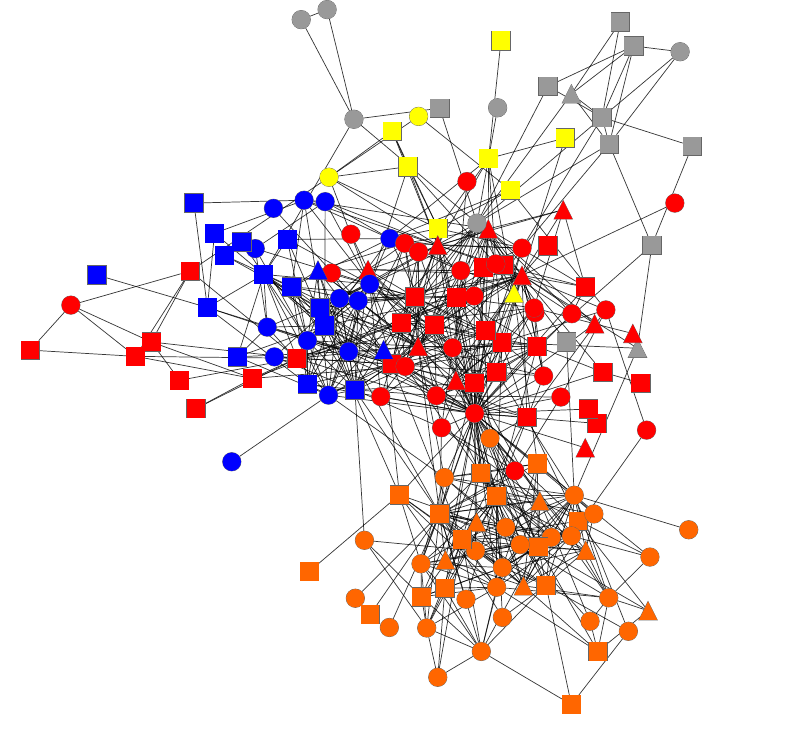}
\end{center}
\caption{
\textbf{S1 Company}: \textit{Blue nodes} - R\&D and administration groups in Asia. \textit{Red nodes} - primarily  hardware verification engineers and chip designers in Asia.
 \textit{Yellow nodes} - Hardware R\&D. \textit{Orange nodes} - acquired startup company. \textit{Gray nodes} - R\&D in Asia. }
\label{S1}
\end{figure}

The S1 company is a publicly held company that specializes in network hardware development. 
According to the company's web page, they employ 500 to 1,000 individuals and have one head office in North America and another in Asia. 
We used the organization crawler to identify 726 informal links among 165 Facebook users who, according to their Facebook page, worked for the company (see Figure~\ref{S1}). 
We also collected information on 84 employee positions inside the company. 
Out of these 84 employees, we identified 20 in management positions.
Most of the discovered company employees held R\&D positions, and most of the identified managers were R\&D team leaders. 

\subsubsection{Small Software Company (S2)}
\begin{figure}[t]
\begin{center}
\includegraphics[
width=0.8\textwidth,clip]{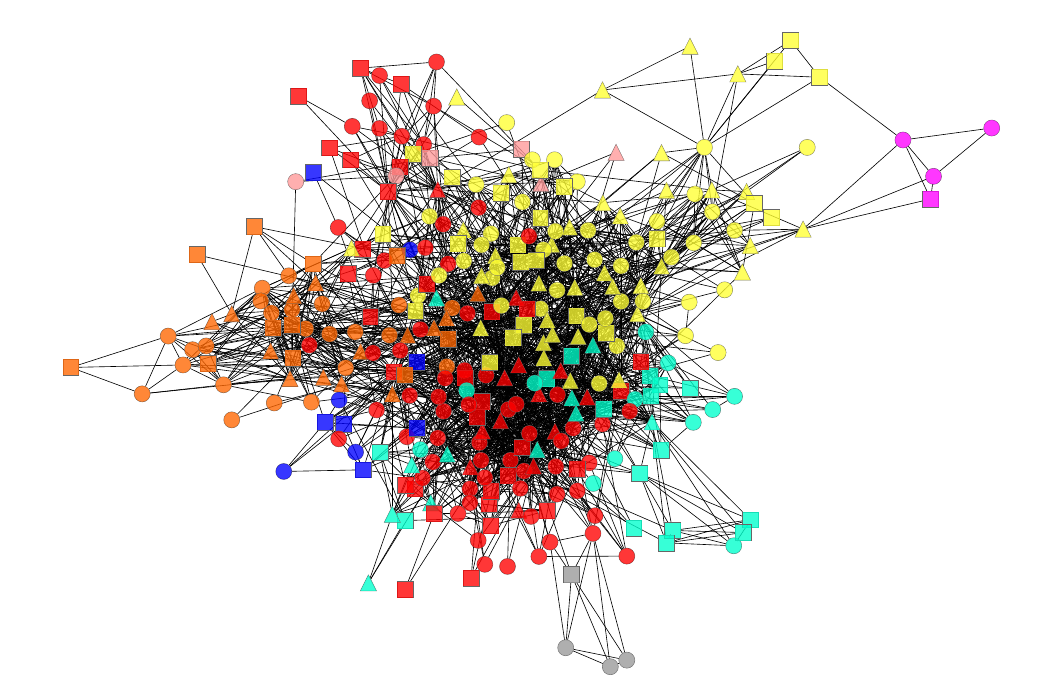}
\end{center}
\caption{
\textbf{S2 Company}: 
\textit{Blue nodes }- IT group in the Middle East. \textit{Red and Orange nodes} - R\&D groups in the Middle East. \textit{Purple nodes} - North American group. \textit{Yellow nodes} - managers and international project managers. \textit{Cyan nodes} - R\&D teams in Australia and the Middle East. \textit{Gray nodes} - European group.
 }
\label{S2}
\end{figure}

The S2 company is an international  publicly held company that specializes in software development. 
According to public sources, the company has between 1,000 and 2,000 employees and maintains offices  in North America, Europe, Asia, Australia, and the Middle East. 
We used our organization crawler to identify 2,369 informal links among 320 Facebook users who stated that they worked for the company in their Facebook profiles (see Figure~\ref{S2}). 
We also collected information on the positions of 168 company employees.
Out of these 168 individuals, 76 were in management positions.
While many of the company employees held project manager (PM) positions, we also identified a number of developers, quality assurance (QA) positions, and support employees.

\subsubsection{Medium Telecommunication Service Company (M1)}
\begin{figure}[t]
\begin{center}
\includegraphics[
width=0.8\textwidth,clip]{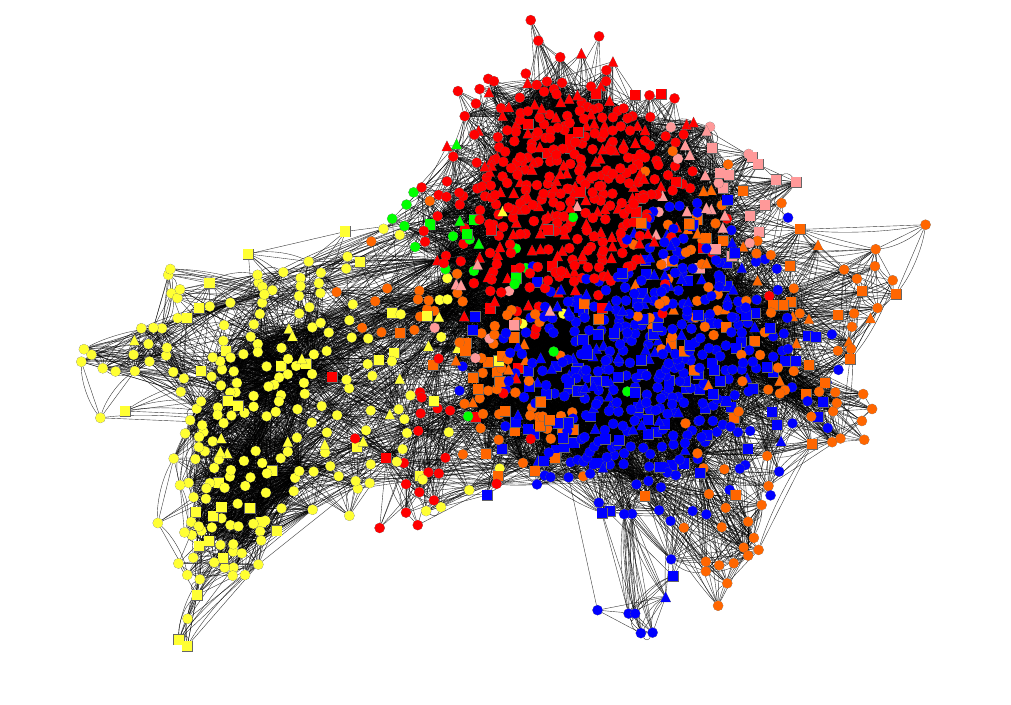}
\end{center}
\caption{
\textbf{M1 Company}: 
\textit{Blue and Orange nodes} - R\&D divisions. \textit{Red nodes} - senior management. \textit{Yellow nodes} - international consultants and support engineers. \textit{Green nodes} - North American headquarter employees.
 }
\label{M1}
\end{figure}

M1 is an international technology company located in North America that specializes in telecommunication services. 
According to the company's web page, M1 currently has between 2,000 and 10,000 employees. 
We used the organization crawler and identified 32,876 informal links among 1,429 Facebook users who, according to their Facebook profile page, worked for the company (see Figure~\ref{M1}).
When we also collected information on the positions of 461 employees,
we learned  227 held management positions.
A wide range of positions inside the company were identified during the crawl: senior management positions, sales and marketing employees, PMs, developers, IT engineers, support engineers, technical writers, etc.

\subsubsection{Medium Software Provider and Outsourcing Company (M2)}
\begin{figure}[t]
\begin{center}
\includegraphics[
width=0.7\textwidth,clip]{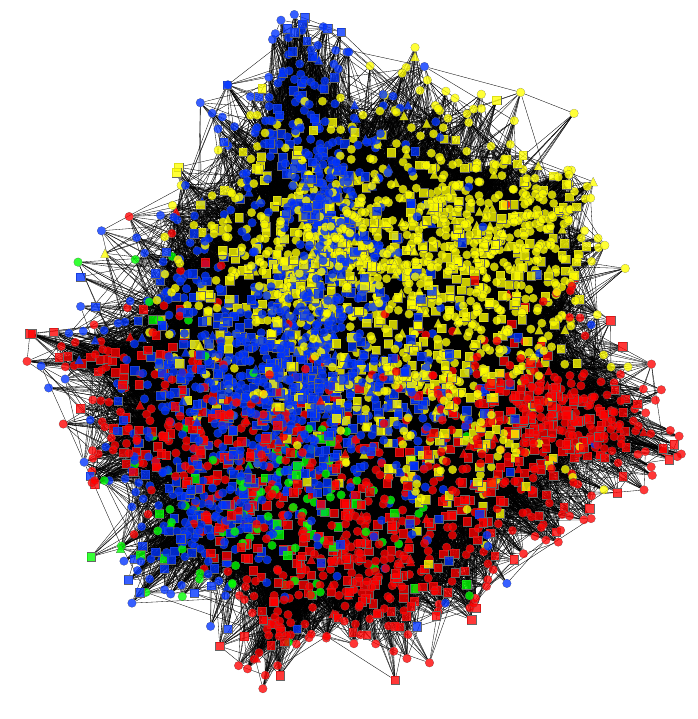}
\end{center}
\caption{
\textbf{M2 Company}: 
\textit{Blue and Green nodes} - R\&D and Specific Domain Experts (SDE) connected to North American and Asia employees. \textit{Red nodes} - R\&D and SDE connected to Australia, Europe and North America. \textit{Yellow nodes} - R\&D and SDE connected to Africa, North America, and Asia. 
 }
\label{M2}
\end{figure}

M2 is an international software and outsourcing provider that specializes in telecommunication services and serves a global customer base. 
The company's web page indicates its size as 10,000 to 20,000 employees. 
We used the organization crawler to focus on the company headquarters, located in South Asia. 
We stopped the crawling process after identifying 87,324 informal links among 3,862 Facebook users who state that they work for M2 in their Facebook profiles (see Figure~\ref{M2}). 
We also succeeded in collecting information on the positions within the company for 1,511 employees. 
During the crawl, a variety of positions were identified: senior managers, developers, sales and marketing positions, IT, PMs, support engineers, technical writers, etc. 
Out of the 1,511 employees, 230 held management positions. 

\subsubsection{Large Information Technology Corporation (L1)}
\begin{figure}[t]
\begin{center}
\includegraphics[
width=0.8\textwidth,clip]{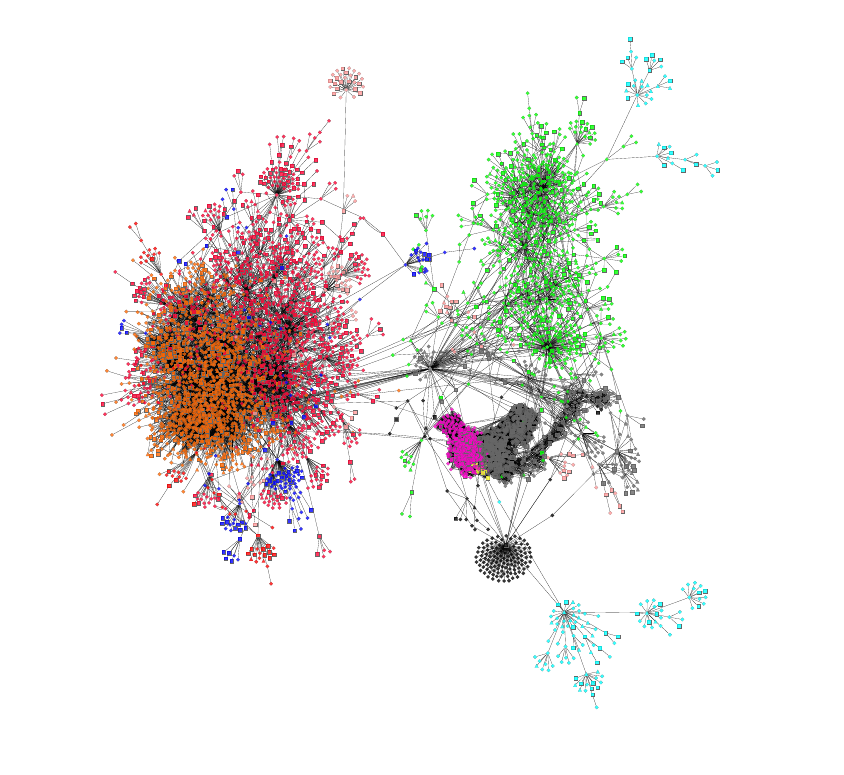}
\end{center}
\caption{
\textbf{L1 Corporation}: 
\textit{Blue nodes} - South American support engineers. \textit{Red nodes} - South American Branch (IT, support engineers, analysts, and PMs). \textit{Orange nodes} - South American Branch (management, sales, marketing, project managers, support engineers, and administration). \textit{Yellow nodes} - Eastern Europe Pricing Analysts. \textit{Purple nodes} - Eastern European consultants 
(marketing, sales and pricing) and support engineers. \textit{Black nodes} - North American Branch, and East Asia - R\&D. \textit{Green nodes} - Middle East R\&D and North American headquarters (management and sales). \textit{Gray nodes} - European consultants and sales and South Asian analysts.
 }
\label{L1}
\end{figure}

L1 is an information technology corporation that provides products and services to customers around the world.
As indicated on the company's web page, L1 currently employs more than 50,000 people.
Our organization crawler collected data on corporation employees in North and South America, Asia, and Eastern Europe.
We identified 45,266 informal links among 5,793 Facebook users who, according to their Facebook profile page, worked for the corporation (see Figure~\ref{L1}). 
We also were able to gather information on the positions of 1,619 employees. 
Out of these 1,619 employees, we succeeded in identifying 463 holding management positions. 
A broad range of positions were identified, spread throughout the world:
senior managers, sales and pricing positions, marketing positions, technical writers, developers, IT, PMs, support engineers, etc.

\subsubsection{Large Technology Corporation (L2)}
\begin{figure}[t]
\begin{center}
\includegraphics[
width=0.8\textwidth,clip]{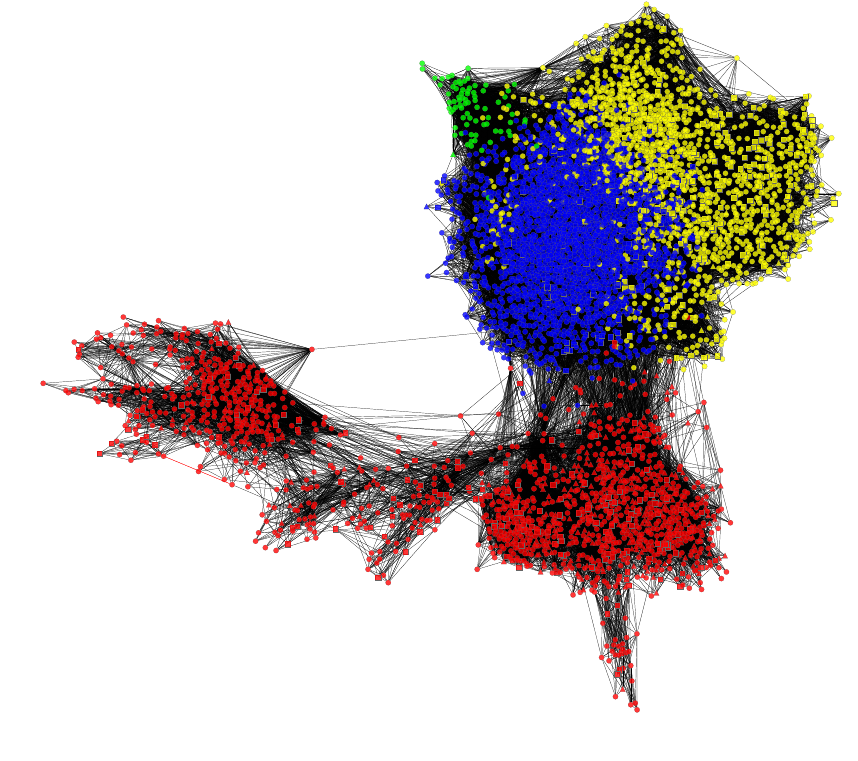}
\end{center}
\caption{
\textbf{L2 Corporation}: 
\textit{Blue nodes} - East Asia Headquarter (management and consultants). \textit{Red nodes} - international senior management and researchers.  \textit{Yellow nodes} - East Asian headquarters (R\&Ds and consultants). \textit{Green nodes} - the company's amateur sports team. }
  
\label{L2}
\end{figure}

The L2 corporation provides hardware and software products, infrastructure, and other technology services to global customers.
According to the company's web page, there are currently more than 50,000 employees.
We used our organization crawler to accumulate data on corporation employees in North and South America, Asia, and Eastern Europe. 
We stopped the crawling process after identifying 94,219 informal links among 5,524 Facebook users who indicated on their Facebook profiles that they worked for the corporation (see Figure~\ref{L2}).
We also succeeded in collecting information on the company positions of 1,131 employees, out of which 461 held management positions.  
During the crawling, we found a wide range of positions inside the company: senior management positions, PMs, sales and marketing positions, developers, IT, support engineers, technical writers, etc.

\begin{table}[ht] 
\begin{center}
\centering % used for centering table 
\caption{Collected Organization Datasets\label{datasets_table}} {
\begin{tabular}{c | l p{2.5cm} l p{2.3cm} } % centered columns (4 columns) 
\hline %inserts double horizontal lines 
  \textbf{Org. }     &	\textbf{Size} &	\textbf{Discovered Employees}& \textbf{Links}  & 	\textbf{Employees Disclosing Positions on Facebook} \\  % inserts table heading   
\hline
 S1 & 500-1K     &	165   & 726	    & 54(32.7\%) \\
\hline
 S2 & 1K-2K   &	320   & 2,369   & 104(32.5\%)\\
\hline
 M1 & 2K-10K   &	1,429 & 32,876	& 383(26.8\%)\\
\hline
 M2 & 10K-20K  &	3,862 & 87,324	&1,531(39.6\%)\\
\hline
 L1 & 50K+	      & 5,793	& 45,266  &1,601(27.6\%)\\
\hline
 L2 & 50K+	      & 5,524	& 94,219  & 1,131(20.5\%) \\
\hline 
\textbf{Total} &        \textbf{-}    & \textbf{17,093}& \textbf{262,780} & \textbf{4,804(28.1\%)}    \\
\hline
\end{tabular} 
}
\end{center}
\end{table}

\section{Identifying Organizational Leadership Roles}
\label{managers}
After the organization crawler completes collecting data from the Facebook profiles of employees of a targeted organization, we can analyze the organizational social network created by the informal Facebook connections. 
In this section, we demonstrate that it is possible to pinpoint leadership roles solely by analyzing the structure of the informal social network of an organization's employees. 

Let $G=<V,E>$ represent the informal social network, where node $v \in V$ is a Facebook user who worked in the target organization and $(u,v) \in E$ represents a Facebook friendship link between two users. 
To pinpoint leadership roles we performed the following steps: First, for each user $v \in V$ in the informal social network, we calculated eight centrality measures.
Next, for each centrality measure, we examined the top 10 and the top 20 users who received the maximal score. 
By reviewing the selected employees' Facebook and LinkedIn profile pages and checking the employment status declared by the individual, we manually classified whether or not the user held a management position (team leader, project manager, vice president, etc.). 
In many cases, however, the user's profile information was not enough to reveal the user's specific position inside the organization. 
To overcome this problem, we cross-referenced the user's personal details with other publicly available online sources, such as Google search engines. By using these methods, in many cases we succeeded in revealing the user's position within the organization.

Lastly, we used several Machine Learning algorithms to build classifiers that can automatically identify management roles inside an organization based on the different centrality measures of the vertices in the informal social network. By using these classifiers, we can recall a wider range of management roles that answer complex centrality measures criteria.
It is important to note that these types of classification methods can be used to compromise users' privacy by exposing non-public positions inside the organization. 
Furthermore, similar methods can assist in revealing various statistics about the organization, thereby disclosing and potentially compromising the organization's privacy.
For example, using the above methods, we estimated the percent of management positions and the number of employees inside each organization (see Tables~\ref{datasets_table} and ~\ref{privacy_table}). 
In many privately held companies, this type of data may be confidential organizational information.

\subsection{Centrality Measures}
Using the organization datasets described Section~\ref{datasets}, we proceeded  to identify leadership roles within the organization using several centrality measures. 
For each node in the informal organization social network, we calculated eight centrality measures:\footnote{The centrality measures were calculated by using the Networkx~\cite{networkx} Python package.} Degree centrality (DG), Closeness centrality (CL)~\cite{newman2005measure}, Betweenness centrality (BC)~\cite{freeman1977set}, HITS (H)~\cite{kleinberg1999authoritative}, PageRank (PR)~\cite{page1999pagerank}, Eigenvector centrality (EC)~\cite{newman2008mathematics}, Communicability centrality (CC)~\cite{estrada2005subgraph}, and Load centrality (LC)~\cite{newman2001scientific}. 
We then sorted the crawled organization's users' list according to the different centrality measures. 

We manually inspected the top 20 user profiles according to each centrality measure in order to infer employees' positions within the target organization.  
Since a large fraction of Facebook users do not disclose their positions on their profile page,  
we used other online sources, such as LinkedIn or results returned by Google's search engine, in order to manually classify whether or not a particular employee held a management position. 
We will refer to managers who do not report their position on Facebook as concealing their management position. 
We determined that the location of an employee in the informal social network of the organization reveals his or her management role within the organization with high precision, even though it was not reported on Facebook. 

Table~\ref{centrality_measures} presents the leadership identification \textit{precision at} the top 10 (T-10) and top 20 (T-20) user profiles for the various centrality measures.  
The results indicate that each of the calculated centrality measures can assist in identifying managers inside the organizations. 
Closeness demonstrated the highest average precision at 20 (0.76), while PageRank received the lowest score (0.70).

\begin{table*}[ht] 
\centering % used for centering table 
\caption{\label{centrality_measures}Management Positions Percentage Based on Centrality Measures (Precision at 10/20)} {% title of Table 
\begin{tabular}{c | c | c c c c c c c c  } % centered columns (4 columns) 
\hline %inserts double horizontal lines 
  \textbf{Org. }     &	\textbf{Cat.}  &\textbf{DG} &	\textbf{CL}& 	\textbf{BC } &  \textbf{H} & \textbf{PR} & \textbf{EC} & \textbf{CC} & \textbf{LC} \\  % inserts table heading 
  \hline
	\multirow{2}{*}{S1}&  T-10          &  0.50             & 0.40              & 0.60         &   0.30      &  0.50      & 0.30           &  0.30    & 0.60     \\
										 &  T-20          &  0.35            & 0.30              & 0.30         &   0.30      &  0.25     & 0.30           &  0.30    & 0.30     \\
	\hline
	\multirow{2}{*}{S2}&  T-10          &  0.80             & 0.90              & 0.80         &   0.90      &  0.70       & 0.90          &  0.90   & 0.80      \\          
	                 	 &  T-20          &  0.70             & 0.75             & 0.75        &   0.70      &  0.75       & 0.70          &  0.75   & 0.75      \\
	\hline
	\multirow{2}{*}{M1}&  T-10         & 1                 & 1                & 0.80        &   1          &  1        & 1               &  1        & 0.80    \\   	
										 &  T-20         & 1                  & 0.95            & 0.85        &   1          &  0.85        & 1               &  1        & 0.85    \\   	
	\hline
	\multirow{2}{*}{M2}& T-10           & 0.83             & 0.71             & 0.86         &   0.83    & 0.86      & 0.83          &  0.83   & 0.88    \\   	
										 & T-20           & 0.73             & 0.82             & 0.69         &   0.80     & 0.71     & 0.80          &  0.80   & 0.69    \\   	
	\hline
	\multirow{2}{*}{L1} & T-10          &  0.55            & 0.80              & 0.80         &   0.78      &  0.60      & 0.78         &  0.78  & 0.80      \\   	
	                    & T-20          &  0.65            & 0.75             & 0.70         &   0.56      &  0.65     & 0.56         &  0.56  & 0.70      \\   	
	\hline
	\multirow{2}{*}{L2} &T-10           &  1               & 1               & 1             &   1          &  1           & 1               &  1        & 1          \\   	
	                    &T-20           &  0.92          & 1               & 1             &   1          &  1           & 1               &  1        & 1          \\   	
\hline
\multirow{2}{*}{\textbf{Avg.}}    &\textbf{T-10} & \textbf{0.78}       & \textbf{0.8}          & \textbf{0.81}       &  \textbf{0.80}    &  \textbf{0.78}     & \textbf{0.80}      &  \textbf{0.80}  & \textbf{0.81}    \\ 
                            &\textbf{T-20} & \textbf{0.725}       & \textbf{0.76}          & \textbf{0.715}       &  \textbf{0.73}    &  \textbf{0.70}     & \textbf{0.73}      &  \textbf{0.735}  & \textbf{0.715}   \\ 
\hline %inserts single line 
\end{tabular} 
}
\end{table*}

Table~\ref{privacy_table} reports the number of concealed management roles that can be detected by using the closeness measure. 
Out of 85 managers detected by focusing on the top 20 Facebook users with the highest Closeness centrality within the informal social network of their organization, 40\% did not report their positions on Facebook.

\begin{table}[ht] 
\begin{center}
\caption{\label{privacy_table}Organization's Hidden Management Positions}{ % title of Table 
\centering % used for centering table 
\begin{tabular}{c | c c c c } % centered columns (4 columns) 
\hline %inserts double horizontal lines 
  \textbf{Org. }     &	\textbf{Classified}  &  \textbf{Classified}         & \textbf{Closeness T20}    &	\textbf{Hidden T20}   \\
  									 &  \textbf{Employee}   &  \textbf{Management}         & \textbf{Management}        &  \textbf{Management}   \\                  
  									 &  \textbf{Positions}  &  \textbf{Positions}   		  & \textbf{Positions}        &  \textbf{Positions}        \\                  
\hline
S1                   & 84                    & 20 (23.80\%)      		        & 6        &  5 (83.33\%)         \\                                       
S2                   & 168                   & 76 (45.20\%)   		          & 15        &  4 (26.66\%)        \\                                       
M1                   & 461                   & 227 (49.20\%)      		        & 19        &  10 (47.40\%)         \\                                       
M2                   & 1,511                 & 223 (14.76\%)   	          & 14        &  2 (14.30\%)        \\                                       
L1                   & 1,619                 & 463 (28.60\%)      		      & 15        &  3 (20\%)         \\                                       
L2                   & 924                   & 461 (49.90\%)   		        & 16        &  10 (62.50\%)        \\                                       
\hline
\textbf{Total}       & \textbf{4,767}	     & \textbf{1,470 (30.80\%)}      & \textbf{85}        &  \textbf{34 (40\%)} \\
\hline 
\end{tabular} 
}
\end{center}
\end{table}

According to these results, high centrality within the informal social network of an organization is a good indication of a leadership role within the organization.  
However, this straightforward general method can only identify management roles of employees with relatively high centrality measures; other management roles with more complex centrality criteria, therefore, will not be identified using this technique. 
To overcome the problem complex centrality criteria,
we used state-of-the art Machine Learning algorithms to classify management roles in each organization (see Section~\ref{ml}).

\subsection{Machine Learning}
\label{ml}
Using state-of-the-art Machine Learning techniques, we constructed classifiers that can identify management positions inside each organization.
This allowed us to identify employees with management roles who satisfied more complex centrality criteria. 
Moreover, using similar methods and techniques assists in distinguishing different types of positions, such as senior management positions and R\&D engineers.

In order to use the Machine Learning algorithm, we first needed to create a training set consisting of sufficient training instances. 
Every training instance represents a collected user in the organization. 
The target attribute is a binary attribute which indicates whether or not the user held a management role inside the organization, while the instance features are the different extracted centrality measures.
We created a sufficient number of training instances by quickly reviewing the users' data extracted from their Facebook profiles.  
By analyzing the crawled organizations' user Facebook profiles, we discovered that an average of 28.1\% of the collected  users had inserted at least partial information about their previous and current employment positions into their Facebook profiles (see Table~\ref{datasets_table}).
For each user who had included his or her previous or current work experience, we attempted to determine if the user held a management role inside the organization. 
In some cases we also did a deeper inspection of the user by cross-referencing the user's work experience with data obtained from other sources, such as LinkedIn. Using this method, we reviewed and classified 4,767 users' profiles.
Out of the 4,767 manually classified positions, we identified 1,470 users who held management positions (see Table~\ref{privacy_table}). 
All these profiles were fed into WEKA~\cite{Hall:2009:WDM:1656274.1656278}, a popular suite of Machine Learning software, as training instances.
Using WEKA, we tested many different Machine Learning algorithms, such as \textit{OneR} (OR), \textit{K-Nearest-Neighbors} (IBk) with $K \in \{1,3,10\}$,  \textit{Naive-Bayes} (NB),  \textit{Decision tree} (J48), \textit{Logistic} (LG), and \textit{RandomForest} (RF). 
Lastly, we evaluated each classifier by using the 10-folds cross validation method and calculating the \textit{Accuracy}, \textit{F-measure}, and \textit{AUC} (Area Under the ROC curve) (see Table~\ref{ml_results}). 
We used T-tests with a significance of 0.05 to compare the different classifiers.  
According to the T-test results, for every organization except S1, all the classifiers returned better accuracy results than the naive \textit{ZeroR} (ZR) classifier. 
Moreover, in most cases the simple OneR classifier is sufficient enough to obtain a near maximum accuracy. However, better AUC and F-measure results were obtained using more advanced classifiers, such as Logistic, and RandomForest classifiers.

\begin{table*}[ht] 
\begin{center}
\caption{Machine Learning Classifiers Results\label{ml_results}}{ % title of Table 
\centering % used for centering table 
\begin{tabular}{c | c |c c| c c c c c c c  } % centered columns (4 columns) 
\hline %inserts double horizontal lines 
  \textbf{Org. }     &	\textbf{Measure}        & \textbf{ZR} &	\textbf{OR}& 	\textbf{J48 } & \textbf{NB}&  \textbf{IBK} & \textbf{IBK} & \textbf{IBK} & \textbf{LG} & \textbf{RF} \\  % inserts table heading 
  							     &								 &								& 								&                  & & \textbf{K=1}& \textbf{K=3}& \textbf{K=10}   & &  \\  % inserts table heading 
  \hline
	\multirow{3}{*}{S1}&  Accuracy        &  \textbf{76.11}        & 68.36        & 71.93         &   72.96      &  65.28 & 74.32      & 75.17           &  73.63      &67.67     \\ 
	                   &  F-measure       &  0           & 0.11           & 0.01         &   \textbf{0.29}      & 0.25      & 0.27           &  0.15    & 0.08 & 0.24     \\
	                   &  AUC             &  0.50         & 0.49           & 0.46         &   0.57      & 0.53      & \textbf{0.64}           &  0.61    & 0.37 & 0.57     \\
	\hline	                   
	\multirow{3}{*}{S2}&  Accuracy        &  54.78       & 60.45          & 62.2         & 63.03        &   63.33      &  61.34      & \textbf{65.13}           &  60.99    & 58.75      \\ 
	                   &  F-measure       &  0           & 0.55           & 0.50         &   0.42      & 0.57      & 0.55           &  \textbf{0.65 }   & 0.48 & 0.55     \\
	                   &  AUC             &  0.50         & 0.60           & 0.60          &\textbf{0.66}         &   0.63      & 0.64      & \textbf{0.66}           &  0.64    & 0.60      \\
  \hline	                   
	\multirow{3}{*}{M1}&  Accuracy        &  50.76       & 65.73          & 66.47        &   63.3      &  61.63      & 67.34           &  65.09    & \textbf{70.72 }&64.67     \\ 
	                   &  F-measure       &  0           & 0.63           & 0.59         &   0.47      & 0.60      & 0.64           &  0.64    & \textbf{0.66} & 0.64     \\
	                   &  AUC             &  0.50         & 0.66           & 0.71         &   0.74      & 0.62      & 0.69           &  0.72    & \textbf{0.76} & 0.69     \\
	
	\hline	                   
	\multirow{3}{*}{M2}&  Accuracy        &  85.24       & 85.13          & 85.96        &   82.24      &  79.14      & 82.69           &  86.45    & \textbf{87 }&83.46     \\ 
	                   &  F-measure       &  0           & 0.22           & 0.22         &   0.32      & 0.26      & 0.26           &  0.29    & \textbf{0.33} & 0.30     \\
	                   &  AUC             &  0.50         & 0.24           & 0.40         &   0.43      & 0.61      & 0.43           &  0.30    & \textbf{0.7 }& 0.58     \\
	
	\hline
	\multirow{3}{*}{L1}&  Accuracy        &  71.4       & 69.15          & 71.61        &   70.79      &  64.40      & 67.36          & 68.43    & \textbf{72.2} &66.28     \\ 
	                   &  F-measure       &  0           & 0.28           & 0.27         &   0.19      & \textbf{0.37}      & 0.34           &  0.33    & 0.11 & \textbf{0.37}     \\
	                   &  AUC             &  0.50         & 0.49           & 0.53         &   0.55      & 0.58      & 0.60           &  \textbf{0.65}    & 0.59 & 0.61     \\
	\hline
	\multirow{3}{*}{L2}&  Accuracy        &  50.22       & 49.91          & 52.92        &   53.9      &  57.11      & 58.53          & 58.66    & \textbf{58.88} &55.71     \\ 
	                   &  F-measure       &  0           & 0.48           & 0.38         &   0.23      & 0.57      & 0.58           &  \textbf{0.61}    & 0.55 & 0.57     \\
	                   &  AUC             &  0.50         & 0.33           & 0.44         &   0.29      & 0.39      & 0.28           &  0.24    & 0.50 & \textbf{0.57}     \\

\hline %inserts single line 
\end{tabular} 
}
\end{center}
\end{table*}

\section{Communities Formed by Employees}
\label{communities}
 \subsection{Community Detection Algorithm}
 \label{communitydetection}

In order to better understand the structure of each organization, we used Cytoscape's Girvan-Newman fast greedy algorithm implementation~\cite{clauset2004finding} to separate each informal social network into disjointed communities.
Each community is marked with a different color in Figures~\ref{S1}-\ref{L2}. 
Node shapes in these figures indicate whether or not the particular employee held a management position in the organization.
Triangle nodes represent those who, to the best of our knowledge, held management positions, while square nodes represent users who did not hold any management position. 
Circles represent employees holding an unknown position within the organization.

\subsection{Community Role Analysis }
After separating the informal social network of each organization into disjoint communities, we analyzed the role of all the major communities within the organization (see Table~\ref{org_communities}).
We cross-referenced the community members with position descriptions and residence locations from their Facebook profile pages. 
We also randomly chose several dozen users from each community.
For these selected users, we manually inspected their positions within the organization by using publicly available sources, such as LinkedIn. 
During this process, we reviewed several thousand employees' profiles and identified the organizational positions of 4,767 users.
The role of each community in the organization was determined by the majority of the community members' positions, geographic locations, and employment histories. 
For example, if most of the sampled community users lived in New York City and worked as software developers within the organization, then we determined that the community was part of the organization's R\&D division in New York City. 
By understanding the role of each community, we inferred details about the organization and the people it employed.
The roles of the different communities within the targeted organizations are presented in Sections~\ref{s1role} -~\ref{l2role}.

\subsubsection{S1 Communities}
\label{s1role}
The community detection algorithm separated the S1 organization social network  into five main communities (see Figure~\ref{S1}).
Community role analysis revealed that S1 has several branches in Asia, most of them consisting of R\&D employees. 
There were four R\&D communities consisting of employees with different sets of skills.
While three communities included mainly software developers (blue, red, and gray communities in Figure~\ref{S1}), one community consisted mainly of hardware developers (yellow community).
Moreover, by reviewing the users' publicly available employment history, we identified a previously acquired start-up company (orange community) and the social connections between the acquired company's employees and S1 employees.

\subsubsection{S2 Communities}
\label{s2role}
The S2 organizational social network was separated into seven communities by the clustering algorithm (see Figure~\ref{S2}).
By reviewing the S2 employees' positions within the organization and user residence locations, we discovered that S2 has one headquarter office in the Middle East (blue, red, and orange communities in Figure~\ref{S2}) and another in North America (purple community). 
We also discovered that the company has worldwide activities occurring on four continents. 
Project managers (yellow community) are living in more than seven different major cities in the world.
The S2 communities' structures indicate that S2 has two headquarters that focus on R\&D and worldwide operations which are managed by the different projects managers in each country.

\subsubsection{M1 Communities}
\label{m1role}
Our clustering algorithm separated the M1 organizational social network graph into five well-connected communities (Figure~\ref{M1}).
We discovered two of the company’s headquarters, both located in North America (green community in Figure~\ref{M1}), and also two large R\&D divisions (blue and orange communities). 
Moreover, we succeeded in detecting the company's senior management community (yellow community) and found the informal connections among the company's senior managers.
Identifying the senior management community may assist in inferring key positions inside the M1 organization that in many cases were not available through publicly available resources.

\subsubsection{M2 Communities}
\label{m2role}
The M2 organizational social network graph was separated into four closely connected communities (Figure~\ref{M2}). 
Each community represents a group of R\&D and Specific Domain Expert (SDE) employees who work in the company's South Asia branch. 
Each one of the four employee groups was well connected to other employee groups within the same company that were located in different parts of the globe. 
For example, the South Asian yellow employees group had close ties with another employee group that was located in Africa, while the red employees group was well connected to employees in Australia, Europe, and North America.

\subsubsection{L1 Communities}
\label{l1role}
Using the community detection algorithm, we separated the L1 social network into 21 communities (Figure~\ref{L1}).
Fourteen of these communities represented nine different roles inside the organization. 
By examining only the residence and position information of these communities, it is possible to pinpoint the group of support engineers in South America (blue community in Figure~\ref{L1}). 
We also succeeded in detecting the company's marketing and sales division in Eastern Europe (yellow and purple communities) and the company's R\&D divisions in North America and East Asia (black community). 
Moreover, we discovered part the company's R\&D group in the Middle East and part of the North American management and sales group (green community).

\subsubsection{L2 Communities}
\label{l2role}
Our community detection algorithm separated the L2 social network into four communities (Figure~\ref{L2}).
Two well-connected communities contain many of the company's R\&D employees, consultants, and managers in the East Asia headquarters (blue and yellow communities in Figure~\ref{L2}). 
We also revealed one of the company's amateur sports teams (green community).
Moreover, we were successful in detecting the corporation's international senior management and their informal connections across four continents and more than 20 major cities (red community).
By analyzing the company's international senior management community, we could discover the cross-Atlantic connections between the different corporate branches.

\begin{table*}[htb] 
\begin{center}
\caption{Organizations' Communities\label{org_communities}}{ % title of Table 
\centering % used for centering table 
\resizebox{\textwidth}{!}{
\begin{tabular}{c | c |c c c c |p{6cm} } % centered columns (4 columns) 
\hline %inserts double horizontal lines 
  \textbf{Org. }     &	\textbf{Comm. } & \textbf{\#Users} &	\textbf{\#Links}& 	\textbf{Number of}          & \textbf{Number of}&\textbf{Description} \\  % inserts table heading   					    
  									 &	       \textbf{Color}                  & &											& 	\textbf{Facebook Profiles}  & \textbf{Classified Users'}&\\  % inserts table heading   							     
  									 &	                         & 											 &											& 	\textbf{with Positions}  	 &  \textbf{Positions}&\\  % inserts table heading   							     
  \hline
	\multirow{11}{*}{S1}&  Blue    								& 30 											& 96 									& 6  													  &  16  &  R\&D and administration groups in Asia\\
	                   \cline{2-7}
	                   &  Red    									& 62											& 234									& 24 														&  37  & Mainly hardware verification engineers and 
	                    																																																									 chip designers in Asia\\
	                   \cline{2-7}
	                   &  Yellow    							& 10											& 13									& 3 														&  8  & Hardware R\&D \\
	                   \cline{2-7}
	                   &  Orange    							& 46											& 197									& 13 														&  21  & Acquired startup company \\
	                   \cline{2-7}
	                   &  Gray       							& 17											& 29									& 5 														&  11  & R\&D in Asia \\	                   
	\hline	                   
	\multirow{14}{*}{S2}&  Blue    								& 10 											& 16 									& 5  													  &  6  &  IT group in the Middle East\\
	                   \cline{2-7}
	                   &  Red    	   	        		& 109											& 645									& 25 														&  45  & R\&D groups in the Middle East \\
	                   \cline{2-7}
	                   &  Orange    	   	        & 48										  & 230									& 16 														&  26  & R\&D groups in the Middle East \\
	                   \cline{2-7}
	                   &  Yellow    							& 100											& 575									& 39 														&  58  & Managers and international PM \\
	                   \cline{2-7}
	                   &  Purple    							& 4	  										& 5									  & 1 														&  1  & Group in North America  \\
	                   \cline{2-7}
	                   &  Gray       							& 4											  & 6									  & 1 														&  1  & European group \\
	                   \cline{2-7}
	                   &  Cyan       							& 39											& 155									& 15 														&  27  & R\&D teams in Australia and the Middle East \\
	                   
	\hline	                   
	\multirow{9}{*}{M1}&  Blue    								& 467 										& 7,685								& 100  													&  163  &  R\&D division\\
	                   \cline{2-7}
	                   &  Red    	   	        		& 425											& 11,706							& 86 														&  129  & Senior management \\
	                   \cline{2-7}
	                   &  Orange    	   	        & 217										  & 2,526							  & 46 														&  75  & R\&D divisions \\
	                   \cline{2-7}
	                   &  Yellow    							& 254											& 3,023							 & 47 														&  51  & International consultants and support engineers \\
	                   \cline{2-7}
	                   &  Green    							  & 23	  								  & 95									& 4 														&  7  & North American Headquarter  \\

	\hline	                   
	\multirow{8}{*}{M2}&  Blue    								& 1,329 										& 23,549					 & 504  													&  498  &  R\&D and SDE connected to North American and Asian 																																																																					 employees\\
	                   \cline{2-7}
	                   &  Red    	   	        		& 1,071											  & 16,637				 &437  												    &  430  & R\&D and SDE connected to Australia, Europe and 																																																																						North America \\
	                   \cline{2-7}
	                   
	                   &  Yellow    							& 1,348											& 24,080					 & 556 														&  551  & R\&D and SDE connected to Africa, North America, 																																																																					 and Asia\\
	                   \cline{2-7}
	                   &  Green    							  & 921	  								  & 1,058									&  33														&  32  & R\&D and SDE connected to North America and Asia 																																																																						employees  \\
  \hline
	\multirow{20}{*}{L1}&  Blue    								& 141 										& 148					 & 45 & 50 &  South America support engineers\\
	                   \cline{2-7}
	                   &  Red    	   	        		& 1,461											  & 1,934				   & 471  												    &  473  & South American Branch (IT, PM, Support engineers, and Analysts) \\
	                   \cline{2-7}
	                   
	                   &  Yellow    							& 15											& 172					 & 6 														&  7  & Eastern European Pricing Analysts\\
	                   \cline{2-7}
	                   &  Orange    							& 1,613											& 7,407					 & 448 											&  422  & South American Branch (Management, Sales, Marketing, PM, Support engineers, and Administration)\\
	                   \cline{2-7}
	                   &  Purple    							  & 443  								  & 13,837						&  100												&  110  & Eastern European (Marketing, Sales and Pricing)  consultants and support engineers \\
	                   \cline{2-7}
	                   &  Green    							  & 921	  								  & 1,482									&  243										&  246  & Middle East R\&D and North American Headquarters (Management and Sales)  \\
	                   \cline{2-7}
	                   &  Gray    							  & 774	  								  & 17,247								&  201												&  175  & European Consultants and Sales and South Asian Analysts. \\
	                   \cline{2-7}
	                   &  Cyan    							  & 154	  								  & 151									&  46														&  67  &East Asian - R\&D \\
	                   \cline{2-7}	                   
	                   &  Black    							  & 143	  								  & 146									&  9														&  9  &North American Branch, East Asian - R\&D \\
	                   \hline
	\multirow{9}{*}{L2}&  Blue    								& 2,285 										& 42,230					 &  220 & 140 &  East Asian Headquarter (management and consultants)\\
	                   \cline{2-7}
	                   &  Red    	   	        		& 1,573											  & 19,841				   & 605  												    &  449  & International Senior management (Senior management, Senior researchers) \\
	                   \cline{2-7}
	                   
	                   &  Yellow    							& 1,588											& 19,023					 &  264														&  218  & East Asian Headquarter (R\&Ds and consultants)\\
	                   \cline{2-7}
	                   &  Green    							& 78											& 1,478					 &  4 											&  1  &The company's amateur sports team\\

  \hline	

\end{tabular} }

}
\end{center}
\end{table*}

\section{Conclusions}
\label{conclusions}
This paper presents methods and algorithms that can be used to collect data from publicly available sources and analyze organizations' social networks.
In order to collect organization datasets, we utilized crawling algorithms based on the homophily principle (see Algorithm~\ref{crawler1}) which can collect organizational data from online social networks like Facebook in matter of hours. 
Using these crawling algorithms, we collected data from the Facebook profiles of employees who worked at six different organizations.
In contrast to the BFS social network crawler, which inefficiently collected organization data, the organization social network crawler presented in this paper succeeded in collecting data from 17,096 social networks users from six organizations with an average precision rate of 22.1\% (Table~\ref{crawler_table}).

We then used the collected organizational data to construct and analyze the informal social network among each organization's employees. 
By calculating eight centrality measures for each selected employee  (i.e., Facebook user) 
we could uncover leadership roles inside the organization (Section~\ref{managers}). 
We discovered that those individuals who received relatively high values in one of the centrality measures were more likely to hold management positions inside the organization. 
Furthermore, the closeness centrality measure presented the best precision at 20 results, with an average precision at 20 of 76\% (see Table~\ref{centrality_measures}). 
Also using the closeness centrality measure, we identified 85 management positions where 40\% of these positions were hidden management roles and did not appear in the individuals' Facebook profiles (see Table~\ref{privacy_table}).

In Section~\ref{ml}, we presented a more sophisticated method for identifying organization management positions by applying Machine Learning algorithms. 
Using WEKA software, we tested and evaluated several algorithms on the datasets which were based solely on calculated centrality measures.
All the evaluated classifiers returned better accuracy results than the trivial ZeroR classifier.
Moreover, better AUC and F-measure results were obtained using more advanced classifiers, such as Logistic, RandomForest, and IBk classifiers (see Table~\ref{ml_results}). 
We believe that these classification results can be improved by adding more features, such as an employee's age and gender as well as the employee's seniority, to the classification algorithm.
Moreover, similar Machine Learning techniques can be applied to identify specific positions inside the organization, such as developers, sales representatives, support engineers, or senior managers.

In this study we also used the community detection algorithm to separate each organization's social network into disjointed communities (see Figures~\ref{S1}-~\ref{L2}). 
By identifying the positions of more than four thousand employees in the organizations studied, we discovered specific community's roles and geographic locations according to the positions and residences of the majority of community users. 
Using this method, we succeeded in inferring many observations about each organization. 
For each organization, we discerned  the geographic locations of its branches and the common employees' qualifications in each branch. 
We also discovered further non-trivial insights about each company. 
For example,  although sample company S1 acquired a start-up R\&D company, the acquired company still performed as a separate company with almost no social connections to S1 as a whole. 
This type of discovery can be used by an organization's management to identify problems within the social structure of the company, such as structural holes~\cite{burt1995structural}.
In the case of company S2, we found this organization  had many project managers who worked in different countries across the world. 
In companies M1 and L2, we uncovered the senior management community and their informal friendship connections. 
Detecting an organization's senior management community can assist in identifying undisclosed management and key positions inside the organization.
Furthermore, by understanding the relationships between a company's senior managers, we can reveal the connections among the organization's different branches.
In the Asian branch of M2, we could infer methods of work where each discovered group inside the Asian branch consisted of R\&D and Specific Domain Expert employees who worked with company's employees in different continents. 
We discovered the L1 company's support divisions in South America and the company's sales and marketing division in Eastern Europe.

We believe this study has several future research directions. 
One possible direction is to create multi-label organizational social networks by cross-referencing an organization's online social network with other social networks associated with that organization, such as the network created by the organization's emails~\cite{tyler}. These  multi-label social networks can provide valuable insights and assist in better understanding the organization as a whole. 
Another possible direction is to combine different community detection algorithms in order to improve an organization's community detection results and reveal more communities inside each organization.
Yet another possible direction is to enrich an organization's user-collected data by automatically adding user data from different publicly available data sources, such as LinkedIn and people search engines. Adding more details to the collected organization's users can improve the results when identifying  community roles within an organization.
A further future direction for this study, which was purposed by Greg Lindsay~\cite{e_nytimes2013}, is to use the collected organizational social network to identify isolated teams inside an organization.

A research direction we have already started to pursue is to examine the implications of malicious users utilizing the collected organizational social network by collect additional information about the organization. Such users might perform a series of friend requests to company employees~\cite{elisharorganizational} or attack a specific employee inside a targeted organization~\cite{hsocialbot}. 

\section{Data Availability}
Anonymous versions of the six organizations' social network topologies used in our study were created by randomly replacing the  employees' Facebook IDs with a series of contiguous integers.  This is available for other researchers to use and can be found on our research group website~\url{http://proj.ise.bgu.ac.il/sns/}. 

% Bibliography
\bibliographystyle{abbrv}
\bibliography{organization_minning}

\end{document}